# Thermally-Limited Current Carrying Ability of Graphene Nanoribbons

Albert D. Liao[1,2,†], Justin Z. Wu[3,4,†], Xinran Wang[4], Kristof Tahy[5], Debdeep Jena[5], Hongjie Dai[4,*], Eric Pop[1,2,6,*]

[1]*Dept. of Electrical & Computer Eng., Univ. Illinois at Urbana-Champaign, Urbana IL 61801*
[2]*Micro and Nanotechnology Lab, Univ. Illinois at Urbana-Champaign, Urbana IL 61801*
[3]*Dept. of Electrical Engineering, Stanford Univ., Stanford, CA 94305*
[4]*Dept. of Chemistry & Laboratory for Advanced Materials, Stanford Univ., Stanford, CA 94305*
[5]*Dept. of Electrical Engineering, University of Notre Dame, Notre Dame, IN 46556, USA*
[6]*Beckman Institute, Univ. Illinois at Urbana-Champaign, Urbana IL 61801*

We investigate high-field transport in graphene nanoribbons (GNRs) on $SiO_2$, up to breakdown. The peak current density is limited by self-heating and scales inversely with width, but can reach >3 mA/μm for GNRs ~15 nm wide. Dissipation from larger devices is limited by the $SiO_2$, but in short GNRs (<~0.3 μm) it also depends on heat flow along the graphene. This allows extraction of a median GNR thermal conductivity (TC), ~80 (130) $Wm^{-1}K^{-1}$ at 20 (600) $^o$C across our samples, dominated by phonons, with estimated <10% electronic contribution. The TC of GNRs is an order of magnitude lower than that of micron-sized graphene on $SiO_2$, suggesting strong roles of edge and defect scattering, and the importance of thermal dissipation in small GNR devices.

[†] These authors contributed equally to this work.

[*] Contact: epop@illinois.edu and hdai@stanford.edu

Graphene nanoribbons (GNRs) are promising materials for nanoelectronics [1, 2], however many unknowns persist about their electrical and thermal properties. Among these, the maximum current density of GNRs is important both for fundamental and practical reasons: it is relevant to know what its limiting mechanisms are, how it compares to carbon nanotubes (CNTs), and to determine the maximum load a GNR transistor could drive within a circuit. By comparison, the current in single-wall CNTs is limited to tens of microamperes in diffusive transport due to self-heating and optical phonon scattering [3, 4], although larger currents can be achieved in short quasi-ballistic samples [5], under ambipolar transport [6], or under avalanche conditions [7]. However, GNRs differ from CNTs in two key aspects: first, they have edges which can cause significant scattering, affecting both electrical and thermal transport [1, 8]. Second, they lie flat on the substrate, which increases their heat dissipation compared to CNTs [9, 10] and can lead to lesser heat-limited current degradation. Nevertheless, to date no studies exist on the maximum current density of GNRs, or their dissipative behavior under high-field transport.

Here, we study the current carrying ability of GNRs on $SiO_2$ up to breakdown, and uncover key roles of heat dissipation both along and perpendicular to the device. We measure current densities up to ~3 mA/μm in ~15 nm wide GNRs, exceeding those typically available in silicon devices. However, the maximum current of GNRs is limited by the high temperature they achieve through Joule self-heating. The high-field behavior and breakdown of GNRs is also sensitive to their thermal conductivity (TC), which enables an extraction of this key parameter.

GNR devices as shown in Fig. 1 were fabricated from solution-deposited GNRs [11], with more details given in the supplement [12]. For comparison, larger exfoliated graphene (XG) samples were also created, with dimensions defined by oxygen plasma pattering. Both types of samples were placed on $SiO_2$ ($t_{ox}$ = 300 nm)/Si substrates, with Si also serving as the back-gate (G). Source (S) and drain (D) electrodes were made with Pd (20 nm) for GNRs and Cr/Au (2/200 nm) for XG devices. GNRs had widths ranging from $W$ = 16-90 nm and lengths $L$ = 0.2-0.7 μm. XG devices had $W$ = 0.1-2.6 μm and lengths $L$ = 3.9-9.7 μm.

To study the upper limits of high-field transport, we measure $I_D$–$V_{DS}$ until devices break from Joule self-heating, as shown in Fig. 1(b). This is similar to the breakdown thermometry technique previously applied to CNTs [10, 13] and nanowires [14]. Like with CNTs, the current drops sharply to zero, creating a small gap in the GNR as imaged in Fig. 1(c). Measurements were made in ambient air, where breakdown (BD) occurs by oxidation at $T_{BD} \approx 600$ °C [10]. By comparison, breakdown of control samples in vacuum (~$7 \times 10^{-6}$ Torr) occurred at six times higher power [Fig. 2(a)], suggesting other failure mechanisms such as defect formation, $SiO_2$ damage [10], or even GNR melting (known to occur at ~3600 °C). An existing graphene model [15, 16] was adapted for GNRs [12], calculating $I_D$ as a function of applied $V_{GS}$, $V_{DS}$ and temperature $T$ under diffusive transport conditions:

$$I_D = qWV_{DS} \left[ \int_0^L \frac{F_x}{n\left(V_{Gx}, T_x\right) \cdot v_d\left(F_x, T_x\right)} dx \right]^{-1} \tag{1}$$

where $q$ is the elementary charge, $x$ is the coordinate along the graphene channel, $n$ is the total carrier density at location $x$, $V_{Gx} = V_G - V_x$ is the potential between gate and location $x$, $F_x = -dV_x/dx$ is the electric field, and $v_d$ is the drift velocity including saturation and temperature effects as in Ref. [16]. The current in eq. (1) is solved self-consistently with the Poisson equation and the heat equation along the GNR [15], both including 3-dimensional (3D) fringing effects in the capacitance [12] and substrate heat dissipation [$g$ in eq. (3) below]. Simulated $I_D$–$V_{DS}$ curves and breakdown voltages in Fig. 1(b) are in good agreement with the experimental data when self-heating (SH) is enabled in the model (solid lines). Without SH the simulated currents are much higher and breakdown is not observed as the temperature remains unchanged.

To gain more physical insight into the scaling of SH in such devices, we consider the power dissipation at breakdown, $P_{BD} = I_{BD}(V_{BD} - I_{BD}R_C)$ [10], where $R_C$ is the electrical contact resistance [12], and $I_{BD}$ and $V_{BD}$ are the current and voltage at breakdown, respectively. We plot $P_{BD}$ vs. the square root of the device channel area in Fig. 2(a). To understand the scaling trend observed, we compare the experimental results with the analytic solution of the heat equation along the graphene devices, similar to CNTs [17]:

$$P_{BD} = gL\left(T_{BD} - T_0\right)\left[ \frac{\cosh\left(L/2L_H\right) + gL_H R_T \sinh\left(L/2L_H\right)}{\cosh\left(L/2L_H\right) + gL_H R_T \sinh\left(L/2L_H\right) - 1} \right] \tag{2}$$

where $T_0$ is the ambient temperature, $L_H$ is the thermal healing length along the device and $g$ is the thermal conductance to substrate per unit length [eq. (3) below]. The thermal resistance at the metal contacts is $R_T = L_{Hm}/[k_m t_m(W + 2L_{Hm})]$. Here $t_m$ is the thickness and $k_m \approx 22$ $Wm^{-1}K^{-1}$ is the TC of the metal electrodes (estimated with the Wiedemann-Franz law [18] using their measured resistivity), and $L_{Hm}$ is the thermal healing length of heat spreading into the metal contacts. The two healing lengths are $L_H = (kWt/g)^{1/2}$ and $L_{Hm} = [k_m/(k_{ox}t_mt_{ox})]^{1/2}$, both of the order ~0.1 μm here. The TC of SiO$_2$ $k_{ox}$ = 1.3 Wm$^{-1}$K$^{-1}$, while $t$ is the thickness and $k$ the TC of the graphene.

The heat loss coefficient into the substrate is written as [12]:

$$g^{-1} = \left\{ \frac{\pi k_{ox}}{\ln\left[6\left(t_{ox}/W + 1\right)\right]} + \frac{k_{ox}}{t_{ox}}W \right\}^{-1} + \frac{R_{Cox}}{W},$$ (3)

and consists of the series combination of the thermal resistance at the graphene/SiO$_2$ interface, $R_{Cox}$ [9, 19, 20], and the 3D spreading thermal resistance into the SiO$_2$ written here as an analytic fit to detailed finite element simulations [12].

The two dashed lines in Fig. 2(a) show the predictions of the model for $k$ = 50 and 500 Wm$^{-1}$K$^{-1}$. We note that for device dimensions $(WL)^{1/2} \gg 0.3$ μm, or approximately three times the healing length, heat dissipation is essentially independent of heat flow along the graphene, and thus on its TC. As a result, dissipation in larger devices made with exfoliated graphene (XG) in Fig. 2 can also be estimated with the simplified approach in Ref. [16]. However for GNRs with dimensions $\leq 3L_H$, heat dissipation occurs in part along the GNR, and this observation is used below to extract their TC. In Fig. 2(b) we plot the maximum current density $I_{BD}$ per width $W$ at the breakdown point (temperature ~$T_{BD}$), and find it can reach over 3 mA/μm for the narrowest GNRs. This current density appears to scale inversely with width which, at first sight, is a counterintuitive finding compared to silicon devices. This also appears at odds with the present understanding that GNRs have significantly lower mobility than large-area graphene [2].

We suggest that GNRs can dissipate more power and thus carry higher current density *at a given temperature* (here, breakdown temperature $T_{BD}$), consistent with a significant role of 3D heat spreading [9]. Figures 3(a-b) display the total device thermal conductance per unit area $G''$ = $P_{BD}/(T_{BD} - T_0)/(WL)$ obtained from the experiments (symbols) and the analytic model from eq. (2) (solid lines). We note that for a given device the maximum power (and current) at breakdown are proportional to $G''$. Similar to Fig. 2(b), we find that both the experimental data and our model scale inversely with the GNR width. To gain a physical understanding of these trends, we consider the heat spreading schematics in Figs. 3(c-e). For 'large' graphene in Fig. 3(c) dissipation occurs mainly 'down' into the oxide. Thus, $G'' = 1/(R_{Cox} + t_{ox}/k_{ox})$ is independent of device dimensions when $L, W \rightarrow \infty$ [in practice $(LW)^{1/2} \gg 3L_H$], as shown with dash-dotted line in Figs.

3(a-b). In general, this expression may include a small heat spreading term into the Si wafer [9, 16], which was negligible here [12]. For large graphene devices the constant expression is also recovered as $G'' = g/W$ when taking the limit $W \to \infty$ of eq. (3).

In contrast, for 'narrow' GNRs the lateral 3D heat spreading into the $SiO_2$ becomes a significant component of the overall thermal conductance of a device [Fig. 3(d)]. In addition, for 'short' devices some heat is conducted along the graphene and into the contacts as well [Fig. 3(e)]. The amount of heat carried out in this manner will depend on the TC and length of the device. The three solid lines in Fig. 3(a) show what the modeled $G''$ predicts for $k = 50, 250,$ and $500$ Wm$^{-1}$K$^{-1}$. As the TC increases, heat is carried more efficiently along the GNR. The device length also matters for 'short' GNRs with $L \leq 3L_H$, when heat generated within the graphene channel is sunk more effectively into the contacts [12, 17]. As a result, the thermal conductance $G''$ increases as $L$ decreases in Fig. 3(b) (also see [12]). In both cases, as the heat dissipation increases, we also see an increase in device current density as plotted in Fig. 2(b), thus confirming that Joule self-heating is a key current limiter in GNR devices.

Since heat dissipation is sensitive to heat flow along 'short' GNRs, it is possible to extract their TC, as shown in Fig. 4. To accomplish this, we iteratively vary $k$ within $L_H$ in eq. (2) until the predicted breakdown power matches the measurements, for each device (we assume a unique $k$ for each GNR). To estimate the confidence intervals of extracted TC for our GNRs, we consider a range $R_{Cox} = 1$–$5 \times 10^{-8}$ m$^2$K/W for the graphene/$SiO_2$ interface thermal resistance [19-21], and an uncertainty of $\pm 1$ layer in the GNR thickness [12]. The extracted TC along with data from the literature on 'large' graphene [22-24] are displayed in Fig. 4(a). We find a TC range $k = 63$-$450$ Wm$^{-1}$K$^{-1}$ for our GNRs, as summarized in the histogram of Fig. 4(b), with a median $\sim 130$ Wm$^{-1}$K$^{-1}$ (at the $T_{BD} = 600$ $^o$C), or $\sim 80$ Wm$^{-1}$K$^{-1}$ at 20 $^o$C, nearly an order of magnitude lower than the TC of exfoliated graphene on $SiO_2$ [23]. The room temperature estimate is done by assuming a mean free path that is independent of temperature (limited by edge or defect scattering), and considering only the temperature variation of graphene heat capacity [10]. Given that we observe no clear dependence of TC on GNR size (i.e. no size effect) in Fig. 4(a), we surmise that here the TC is limited by edge roughness and defect or impurity scattering. However, the *range* of values extracted can be attributed to *variations* in edge roughness and defect or impurity density between samples [12]. For instance, recent scanning tunneling microscopy (STM) studies [25] have found that edges of such GNRs vary from atomically smooth to $\sim 1$ nm edge roughness. Simulations [8, 26] suggest that edge disorder could nearly account for the variation in TC observed in Fig. 4, while different impurity or defect density between samples will only serve to broaden the observed distribution.

Before concluding, we examine if the thermal and electrical properties of the GNRs are related, and plot the extracted TC vs. the inverse sheet resistance in the Fig. 4(b) inset. Also plotted is the electronic contribution to TC ($k_e$), estimated with the Wiedemann-Franz law [18] to be nearly always an order magnitude lower ($< 10$ Wm$^{-1}$K$^{-1}$). This estimate is likely an upper limit, as the Lorenz number ($L_0 = 2.45{\times}10^{-8}$ W$\Omega$K$^{-2}$) in graphite is unchanged [27], but in nanostructures where edge scattering dominates it is slightly lower than the bulk value [28]. This simple analysis suggests that TC of GNRs is dominated by phonons at room temperature and above. However, TC and electrical conductance follow similar trends here, indicating that similar scattering mechanisms limit both phonon and electron transport. These scatterers include edges, impurities and defects in GNRs [2, 8, 26, 29].

In conclusion, we have shown that high-field transport in GNRs on SiO$_2$ is limited by self-heating. The maximum current density at a given temperature scales inversely with GNR width and reaches >3 mA/μm in ~15 nm wide devices. Dissipation in 'large' graphene ($\gg$0.3 μm, or three times the thermal healing length) is limited primarily by the SiO$_2$ thickness, but dissipation in 'small' GNRs improves from 3D heat spreading into the SiO$_2$ and heat flow along the GNR to the contacts. Taking advantage of this sensitivity we found a median TC ~ 80 Wm$^{-1}$K$^{-1}$ for GNRs at room temperature, with less than 10% electronic contribution. We acknowledge funding from the Marco MSD Center, the ONR, AFOSR (EP), the NRI and the NRI Coufal fellowship (AL).

**Figure captions:**

**FIG. 1** (color online). (a) Schematic of graphene devices used in this work. (b) Measured (symbols) and simulated (lines) current-voltage up to breakdown of GNRs in air. Solid lines are model with self-heating (SH) and breakdown when max($T$) > $T_{BD}$ = 873 K, dashed lines are isothermal model without SH (see text and supplement [12]). Dimensions are $L/W$ = 510/20 nm for D1, and $L/W$ = 390/38 nm for D2. $V_{GS}$ = -40 V to limit hysteresis effects. (c) Atomic force microscopy (AFM) image of D1 after high-current sweep; arrow shows breakdown location.

**FIG. 2** (color online). (a) Scaling of GNR and XG breakdown power with square root of device footprint, $(WL)^{1/2}$. Dashed lines are thermal model with $k$ = 50 and 500 Wm$^{-1}$K$^{-1}$, $R_{Cox}$ = 5×10$^{-8}$ m$^2$KW$^{-1}$ and $L/W$ = 15. Lateral heat sinking and in-plane GNR thermal conductivity begin to play a role in devices < ~0.3 µm (also see Fig. 3). A few devices were broken in vacuum as a control group. (b) Scaling of peak current at breakdown vs. device width, demonstrating greater current density in narrower GNRs that benefit from 3D heat spreading and lateral heat flow along the GNR. Dashed line drawn to guide the eye.

**FIG. 3** (color online). Thermal conductance of device per unit area ($G''$) vs. width for graphene of varying (a) thermal conductivity and (b) length. Both parameters affect heat sinking along GNRs < ~0.3 µm. Symbols follow the notation of Fig. 2. Horizontal dash-dotted line is the limit $W \rightarrow \infty$ which applies to the case in (c), only "vertical" heat sinking through the oxide. The significance of lateral heat spreading from GNRs is shown in (d) and (e), both mechanisms partly leading to higher current density in Fig. 2.

**FIG. 4** (color online). (a) Thermal conductivity (TC) of GNRs from this work, compared to large-area graphene measurements from the literature [22-24]. (b) Histogram of TC for our GNRs shows a range 63–450 Wm$^{-1}$K$^{-1}$ with a median of 130 Wm$^{-1}$K$^{-1}$ at the breakdown temperature (600 $^o$C). The median TC at room temperature is ~40% lower, or ~80 Wm$^{-1}$K$^{-1}$, nearly an order of magnitude lower than 'large' exfoliated graphene (XG) on SiO$_2$ [23]. The inset shows approximate scaling between TC and electrical sheet conductance, suggesting scattering mechanisms common to both electrons and phonons. The electronic contribution to thermal conductivity ($k_e$) is estimated with the Wiedemann-Franz law to be typically <10 Wm$^{-1}$K$^{-1}$ or <10%. Dashed lines show trends to guide the eye.

Figure 1:

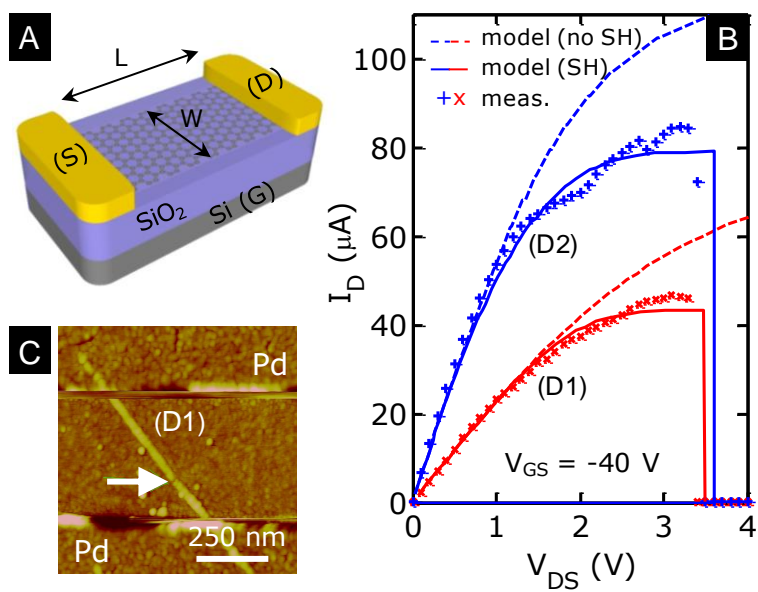

Figure 2:

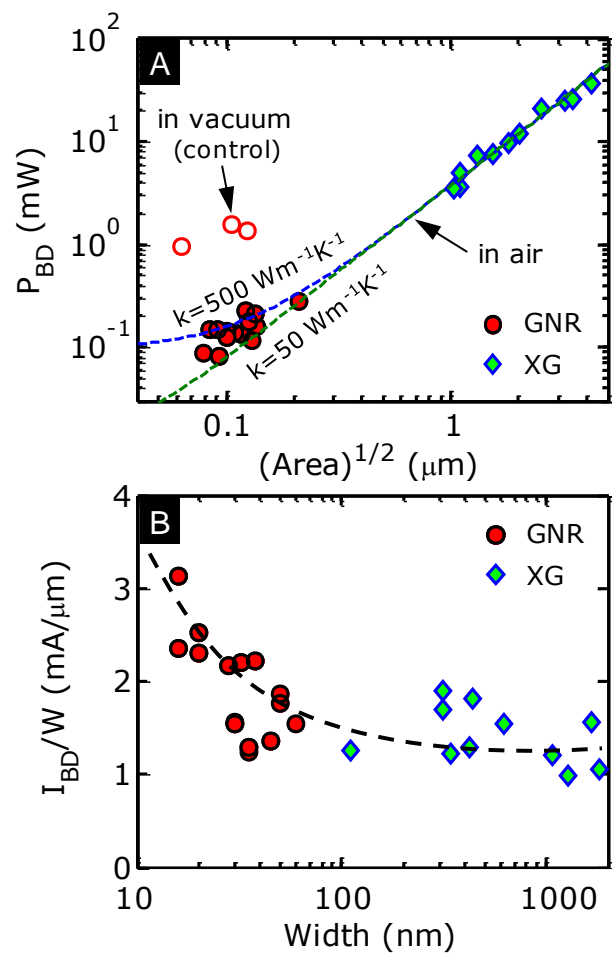

**Figure 3:**

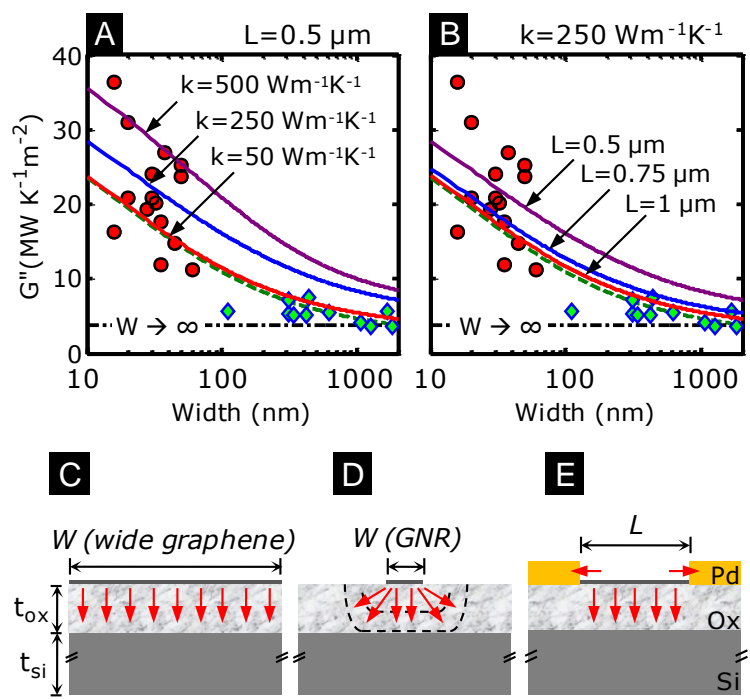

Figure 4:

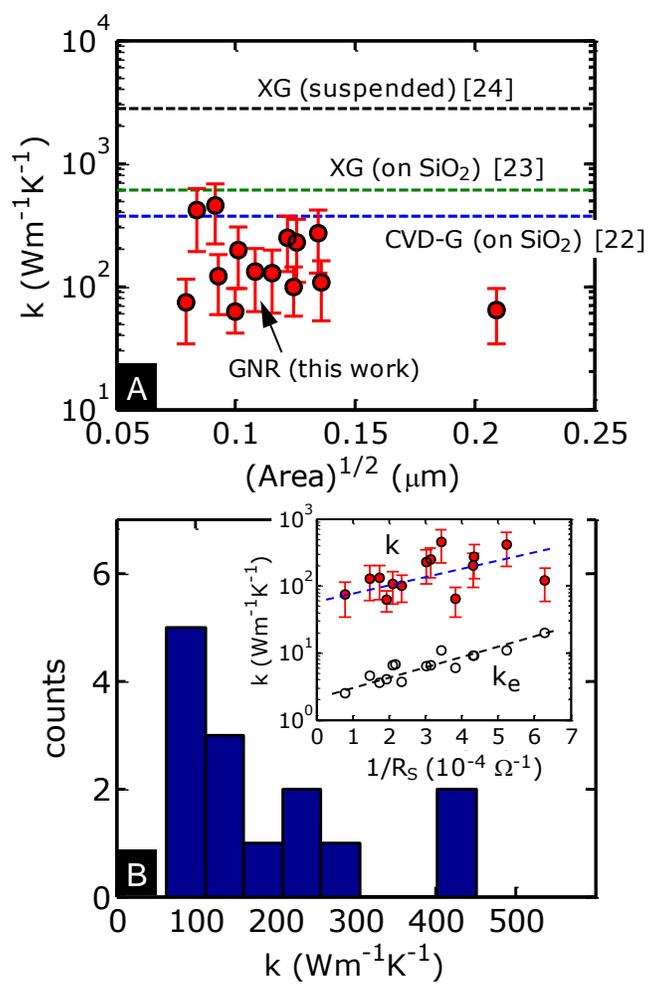

# Supplementary Information

for "Thermally-Limited Current Carrying Ability of Graphene Nanoribbons" by Albert Liao, Justin Wu, Xinran Wang, Kristof Tahy, Debdeep Jena, Hongjie Dai and Eric Pop

## 1. Fabrication Details

Graphene nanoribbon (GNR) devices were obtained from a 1,2-dichloroethane organic solution of poly(m-phenylenevinylene-co-2,5-dioctoxy-p-phenylenevinylene) (PmPV) by sonication of pristine multi-wall nanotubes (MWNTs) that had been calcined at 650 $^{o}$C [1]. An ultracentrifuge step was performed to remove the remaining nanotubes, following the method described by Jiao et al [1]. The solution was spin coated onto ~300 nm SiO$_2$ substrates on highly doped silicon wafers. After calcination of the coated substrate at 275 $^{o}$C for 20 minutes to remove the remaining PmPV, an array of 20 nm thick Pd electrodes and pads were defined by e-beam evaporation and lift-off. The contacts were annealed in Ar at 200 $^{o}$C after which the devices were then annealed electrically. Samples were characterized by atomic force microscopy (AFM) and electrical testing to determine which of electrode pairs correspond to valid devices. Such devices consist primarily of non-AB stacked 2-layer GNRs, however some layer variation is seen in Fig. S1 and is characterized in Section 2 below.

For comparison, micron-sized exfoliated graphene (XG) devices from Graphene Industries were deposited on heavily n-type doped silicon wafers, also with ~300 nm thermal oxide. XG devices were identified using optical and Raman microscopy. The wafers were backside-metalized after oxide removal in HF to form back-gate contacts. The graphene flakes were then patterned using an O$_2$ plasma reactive ion etch with PMMA masks. E-beam evaporated Cr/Au (2/200 nm) was used to define the drain and source contacts by e-beam lithography. After metal deposition and lift-off, the samples were annealed in forming gas at ~400 °C for ~2 hours to remove the e-beam resist residue.

## 2. Thickness of GNRs:

As the graphene nanoribbons (GNRs) are fabricated from multi-walled carbon nanotubes [1], the layer stacking orientation is random unlike bulk graphite. Thus we cannot use Raman spectroscopy to count the number of layers. Instead we rely on the measured thickness from AFM scans to distinguish the number of layers of graphene, as shown in Figure S1. We note that numerical values from AFM scans (Fig. S1) are used only for counting layers and not in our calculations, because they are not an accurate measurement of the real thickness. The actual thickness that is used in calculations is the number of layers times the inter-atomic spacing between graphene sheets (0.34 nm). We also note that AFM images of ribbons and sheets show similar heights for similar layer numbers.

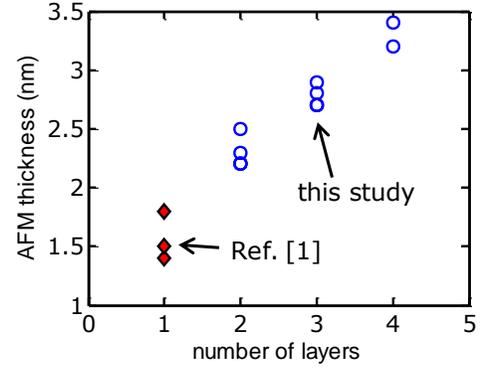

**Figure S1.** Measured AFM thickness of GNR samples vs. number of layers assigned. GNRs that correspond to this study are represented with open circles and ones from Ref. [1] in filled diamonds.

### 3. Contact Resistance:

We subtract the power dissipated at the contacts ($I^2 R_C$, where $R_C$ is the contact resistance) from the total power measured, to obtain only the power dissipated within the graphene. Because the GNR test structures had two terminals, an indirect method of extracting $R_C$ was employed. Measuring the low bias (LB) resistance, which depends geometrically on the sheet and contact resistance [2], we fit the following expression to our data:

$$R_{LB} = \frac{\rho_C}{W} + \frac{\rho_S L}{Wt} \ . \tag{E1}$$

Here $R_{LB}$ is the low bias resistance taken from the linear region of the $I_D$-$V_{DS}$ measurement, $\rho_C$ is the contact resistivity, $\rho_S$ is the sheet resistivity of graphene, $L$ is the length, $W$ is the width, and $t$ is the thickness of the sample. All measurements were performed at $V_{GS}$ = -40 V back-gate bias, which eliminates hysteresis effects. We find that variations arising from fabrication have a notable impact on the values extracted when comparing two different batches of devices shown in Fig. S2. A fit to the data from a given batch is made using eq. E1 (dashed lines) and an average contact resistance is extracted. The average contact resistivity for batch 1 is $\rho_C = R_C W$ ~ 630 Ω·μm and for batch 2 $\rho_C = R_C W$ ~ 250 Ω·μm, which are both typical for graphene-Pd contacts, and within experimental variation observed by us and other groups. From this, we also obtain the sheet resistance ($\rho_S/t$) for the inset of Fig. 4(b) in the main text.

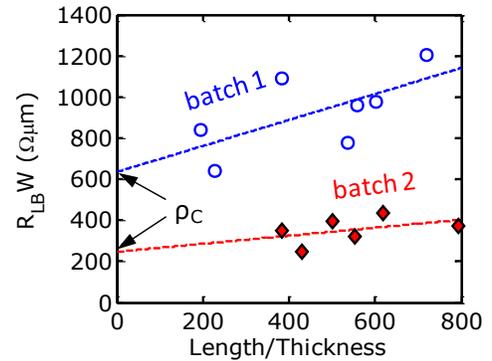

**Figure S2.** Measured low-bias resistance ($R_{LB}$) times width vs. length/thickness ratio of the GNRs. The y-intercept is the average contact resistance times width. Two fabrication batches yielded different contacts.

## 4. Spreading Thermal Resistance of GNRs:

We have calculated the thermal spreading resistance of GNRs on SiO$_2$ by finite element (FE) simulations, and then obtained a simple analytic model that matches the FE calculations over a wide range of parameters. We note that for 'large' graphene devices of dimensions $(WL)^{1/2} \gg 3L_H$ (typically $\gg$ 0.3 µm) the heat sinking occurs simply 'vertically' through the underlying SiO$_2$, and a simple expression for the overall thermal resistance given by Ref. [3] holds, $R_{th} = 1/(hA) + t_{ox}/(k_{ox}A) + 1/(2k_{Si}A^{1/2})$ with $A = LW$ the area of the channel, $h \approx 10^8$ Wm$^{-2}$K$^{-1}$ the thermal conductance of the graphene-SiO$_2$ boundary [4], $k_{ox}$ and $k_{Si}$ the thermal conductivities of SiO$_2$ and of the doped Si wafer. However, for GNRs there is significant *lateral* heat spreading into the SiO$_2$ and *along* the GNR. In Fig. S3 we show results of a few FE simulations of GNRs dissipating heat into SiO$_2$; it is clear that a simple model of vertical heat flow is inadequate.

We then define the heat dissipation coefficient from GNR into substrate per unit length ($g$). To understand its physical meaning, consider that in steady-state, for a *long* GNR (far from the contacts), the temperature at location $x$ along the ribbon is simply given by $T(x) = T_0 + p'(x)/g$ where $p'(x)$ is the power dissipated per unit length (in units of W/m) at location $x$ [5]. We numerically calculate the coefficient $g$ for a variety of graphene device widths, from 10 nm to 6 µm and show the results as symbols in Fig. S4, both for oxide thickness t$_{ox}$ = 90 and 300 nm. Here the lines are the results of a fit with the simple expression:

$$g^{-1} = \left\{ \frac{\pi k_{ox}}{\ln\left[6(t_{ox}/W + 1)\right]} + \frac{k_{ox}}{t_{ox}}W \right\}^{-1} + \frac{R_{Cox}}{W}, \qquad (E2)$$

which (as in the main text) consists of the series combination of the thermal boundary resistance at the graphene/SiO$_2$ interface, $R_{Cox}$ (see main text) and the spreading thermal resistance into the SiO$_2$ written here as an analytic fit to the detailed FE simulations. The simple expression applies to both wide and narrow graphene ribbons, and includes both the graphene/SiO$_2$ interface and the thermal conductivity of the SiO$_2$. The thermal conductivity of the underlying Si wafer is not included as it was found (from the FE simulations) that it plays a very minimal role.

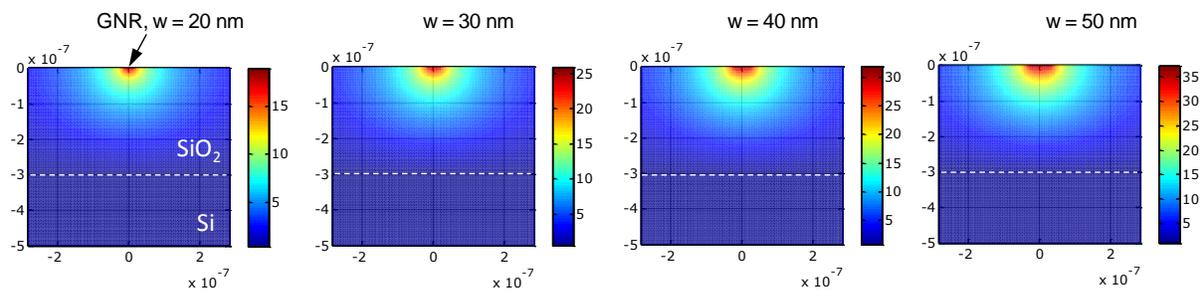

**Figure S3.** Finite-element (FE) simulation cross-section of heat spreading from GNRs with width from W = 20-50 nm on SiO$_2$ with thickness t$_{ox}$ = 300 nm.

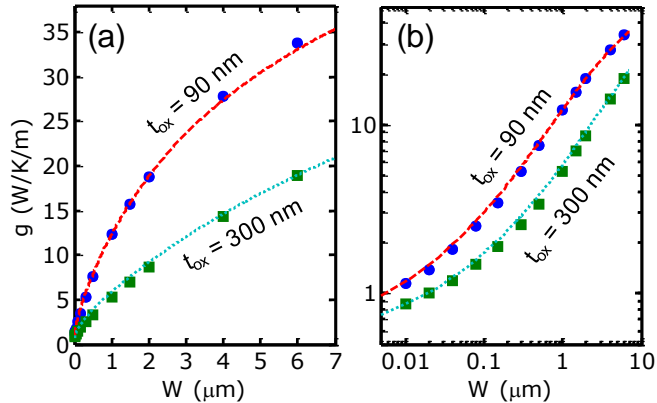

**Figure S4.** Heat spreading coefficient per unit length from graphene device into substrate, for SiO$_2$ thicknesses, t$_{ox}$ = 90 and 300 nm. Analytic model from eq. E2 (lines) shows excellent agreement with finite-element simulations (symbols). Panels (a) and (b) display the same results, with linear and logarithmic scales.

Note that by comparison, the thermal dissipation per length from carbon nanotubes (CNTs) on SiO$_2$ is of the order $g_{CNT}$ ~ 0.2 WK$^{-1}$m$^{-1}$ [5] and thus much lower than in GNRs which are always wider. This value for CNTs was incorrectly applied to GNRs in a previous study [6], leading to an over-estimate of the thermal conductivity of GNRs.

## 5. Graphene Model Extended to GNRs:

To obtain the current-voltage calculations displayed in Fig. 1(b), we extended a previously developed graphene finite-element simulation [3, 7]. Modifications include a lower mobility, as is typical of such GNRs [8], $\mu_{0,1}$ = 160 cm$^2$V$^{-1}$s$^{-1}$ for D1 and $\mu_{0,2}$ = 280 cm$^2$V$^{-1}$s$^{-1}$ for D2. A contact resistance $R_C W$ = 250 Ω·μm was used for both devices in accord with Section #1 above. Together, these parameters were sufficient to model the low-field behavior in Fig. 1(b).

To model the high-field and temperature-dependent behavior we used the velocity saturation model in [3]. The temperature dependence of mobility used was $\mu(T) = \mu_0(300/T)^{1/2}$, similar to carbon nanotubes [9], but slightly weaker than in 'large' graphene on SiO$_2$ [3]. We note that decisive temperature-dependent mobility data are not yet available for GNRs, and it is likely these would change from sample to sample due to variation in impurity and edge scattering. However, weaker temperature dependence of mobility in GNRs is reasonably expected, as similarly observed in metal nanowires vs. bulk metals [10]. Regardless, as it turns out, the specific mobility model has less impact on the high-field behavior of GNRs, which is dominated by the high-field saturation velocity, including its carrier density and temperature dependence [3].

To calculate the charge density along the GNR we use the approach in Ref. [3], but here we must include fringing effects in the capacitance of the GNR above the Si back-gate. The GNR capacitance per unit area is modeled similarly to the spreading heat effect in Section #3 above:

$$C_{ox} = \varepsilon_{ox}\varepsilon_0\left\{\frac{\pi}{\ln\left[6(t_{ox}/W+1)\right]W} + \frac{1}{t_{ox}}\right\}, \tag{E3}$$

where the first term represents the fringing capacitance and the second term is the parallel plate capacitance of the GNR. The expression reduces to the familiar $C_{ox} = \epsilon_{ox}\epsilon_0/t_{ox}$ (in Farads per unit area) in the limit $W \to \infty$ as expected for large graphene, and is in good agreement with finite-element simulations [11].

The temperature along the GNR was computed iteratively with the finite-element method described in Ref. [7], using $k_1 = 100$ Wm$^{-1}$K$^{-1}$ and $k_2 = 175$ Wm$^{-1}$K$^{-1}$, scaled consistently with the mobility of the two GNRs [note Fig. 4(b) inset in main text], and consistent with thermal conductivity values extracted in the overall study. The heat loss coefficient from GNR to substrate was modeled using the heat spreading expression from eq. E2 above.

## 6. Length Dependence on Total Thermal Conductance:

The length dependence on the total thermal conductance $G''$ is similar to that between $G''$ and the width as shown in Fig S5. However, the reason for the inverse relation with length is different. As the length of the GNR decreases to the point where it is on the order of a few healing lengths ($L_H$), more heat can be dissipated into the contacts. It is this dependence on length that allows us to extract the thermal conductivity of our samples as indicated in Fig. 4 and Fig. S5.

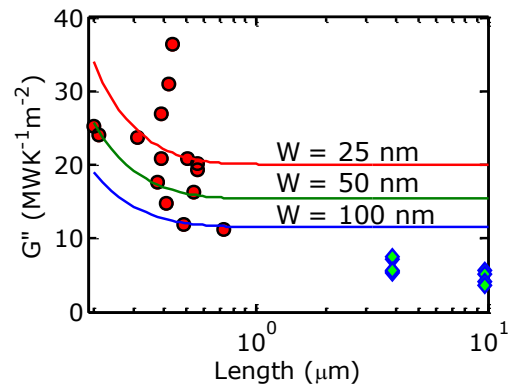

**Figure S5.** Length dependence on $G''$ plotted for 3 different widths, with k = 100 Wm$^{-1}$K$^{-1}$. The length of most of our GNR samples falls in the range where G'' is sensitive to k, i.e. ~3 healing lengths long.

## Supplementary References: